\documentstyle[12pt]{article}
\normalbaselines
\begin{document}
\bibliographystyle{unsrt}
\pagenumbering {arabic}
\vbox{\vspace{38mm}}
\begin{center}
{\LARGE \bf W GRAVITY, $N=2$ STRINGS AND \\[2mm]
 $ 2+2~SU^*(\infty)$ YANG-MILLS INSTANTONS  }\\[5 mm]

Carlos Castro \\{I.A.E.C 1407 Alegria\\
Austin, Texas 78757  USA }\\[3mm]
(August  1993)\\[5mm]
\end{center}

\begin{abstract}

     We conjecture that $W$ gravity can be interpreted as the gauge theory
of  $\phi$ diffeomorphisms in the space  of dimensionally-reduced $D=2+2$
$SU^*(\infty)$  Yang-Mills instantons. These $\phi$
diffeomorphisms preserve a volume-three form and are those which furnish  the
correspondence between
the dimensionally-reduced Plebanski equation and the KP equation in $(1+2)$
dimensions. A supersymmetric extension furnishes  super-$W$ gravity.
The Super-Plebanski equation  generates  self-dual
complexified super gravitational backgrounds (SDSG) in terms of the
super-Plebanski  second heavenly form. Since the latter
equation yields  $N=1~D=4~SDSG$ complexified  backgrounds  associated
with the complexified-cotangent space of the Riemannian surface,
 $(T^*\Sigma)^c$,  required in the formulation of $SU^*(\infty)$
complexified Self-Dual Yang-Mills theory, (SDYM );  it naturally follows that
the
recently constructed $D=2+2~N=4$ SDSYM  theory- as the consistent background of
the open $N=2$ superstring- can be embedded into the
$N=1~SU^*(\infty)$ complexified Self-Dual-Super-Yang-Mills (SDSYM)  in $D=3+3$
dimensions. This is achieved after using  a generalization of self-duality for
$D>4$.  We finally comment on the the plausible relationship between
the geometry  of $N=2$ strings and the moduli  of $SU^*(\infty)$
complexified  SDSYM in $3+3$ dimensions.
\end {abstract}

\section {Introduction}

  Recently [1,2], starting from a self-dual $SU^*(\infty)$ (complexified)
Yang-Mills theory
in  $(2+2)$
dimensions, the Plebanski second heavenly equation was obtained after a
suitable dimensional reduction. The self-dual gravitational background was
the complexified  cotangent space of the internal two-dimensional Riemannian
surface
required in the formulation of $SU^*(\infty)$ Yang-Mills theory [22]. A
subsequent
dimensional reduction leads to the  KP equation in $(1+2)$ dimensions after
the relationship from the Plebanski second heavenly function, $\Omega$, to
the KP function, $u$, was obtained. Also a complexified KP equation
was found when a different dimensional reduction scheme was performed.  Such
relationship between $\Omega$ and $u$  is based on the
correspondence between the $SL(2,R)$ self-duality conditions in $(3+3)$
dimensions of Das, Khviengia, Sezgin (DKS) [3] and the ones of $SU(\infty)$
 in $(2+2)$ dimensions . The generalization to the
Supersymmetric KP equation is  straightforward by
extending the construction of the bosonic case to the previous Super-Plebanski
equation, found by us in
[1], yielding  self-dual supergravity backgrounds
in terms of the light-cone  chiral superfield, $\Theta$, which is the
supersymmetric analog of $\Omega$. The most important consequence of this
Plebanski-KP correspondence is that $W$ gravity can be seen as the gauge
theory of  $\phi$-diffeomorphisms  in the space of dimensionally-reduced
$D=2+2,~SU^*(\infty)$ Yang-Mills instantons. These $\phi$ diffeomorphisms
preserve a volume-three-form and are, precisely, the ones which provide the
Plebanski-KP correspondence. This was the main result of [2].

     As a byproduct, we can generalize now  our results to the supersymmetric
case where
the super KP equation can be obtained from the Super-Plebanski one using
the results of [1]. The supersymmetric case is  very relevant because there
has been a lot of work recently [4] pertaining to the crucial role that $N=4$
SDSYM in $D=2+2$ has as the consistent $N=2$ open superstring background
[5]. The relevance of $N=8$ (gauged) Self-Dual-Supergravity (SDSG)   for
the consistent background of the  closed $N=2$ (heterotic)
superstring was noticed by Siegel [5]. Lately Nishino has shown [4] that the
$N=4$
SDSYM in $D=2+2$ generates Witten's topological field theory after
appropriate twisted dimensional reduction/truncations. Therefore, the
supersymmetric case is very relevant to current research.

    There has been many derivations of the (super) KP equation in the
literature. In [4] Supersymmetric KP systems were embedded in a  SDSYM
theory in $D=2+2$. DKS [3] derived the KP equation from a self-duality
condition in $(3+3)$ dimensions. Barcelos-Neto et al [7] derived it from
the vanishing curvature equation in $2+1$ dimensions for an $SL(2,R)$
valued potential.  A derivation of the KP equation based on the asymptotic
$h\rightarrow 0$
limit of the continual $sl(N+1,C)$ Toda molecule equation was given earlier by
Chakravarty and Ablowitz [8]. The continual Toda molecule equation was
obtained,  first,  by a suitable ansatz, dimensional reduction and continuous
version of the the Cartan basis for $sl(N+1,C)$. And many others. Six
reasons why our results in [2] differed from the previous authors were
outlined in [2]. The seven, and most important difference, was that we were
able to provide with the geometrical origin of $W$ (super) gravity !
The former can be interpreted  as a gauge theory in the
space of $D=2+2~SU^*(\infty)$ (super) Yang-Mills instantons !

     The outline of this letter  goes as follows : In section $II$ we review
the derivation in [2] and show why $W$ gravity can be interpreted as the
gauge theory of $\phi$ diffeomorphisms in the space of dimensionally
reduced $D=2+2~SU^*(\infty)$ Yang-Mills instantons. The $W~ metric$ is the
gauge field that gauges these $\phi$ volume-preserving diffs. We discuss a
very natural and plausible  procedure to  $truncate$ the theory and  yield
$W_N$ gravity. A  way how to obtain $W_{\infty}(\lambda)$ algebras is also
discussed briefly. Finally, in $III$ we
present a  discussion about  the supersymmetric case and point out  why the
importance of  $N=4~SDSYM$
 in $D=2+2$- as a consistent $N=2$ open superstring background-  really
originates
from a dimensional reduction of the $N=1~SU^*(\infty)~SDSYM$ in $D=3+3$. (
Of course, for those Lie algebras that can be embedded in $SU^*(\infty)$).
i.e; the geometry (moduli) of the $N=2$ open superstring stems from the
moduli space of $N=1~D=3+3~SU^*(\infty)$ SDSYM theory. This is not
surprising in view of the ubiquitous presence of Calabi-Yau manifolds
in the string literature. Generalized self-duality conditions for spaces
with $D>4$ are currently under investigation by many people [23].

    \section {On $W$ gravity and the KP equation from Plebanski }

    Let us choose complex coordinates for the complexified-
spacetime, $C^4$;  $y=(1/\sqrt 2)(x_1+ix_2);\tilde
y=(1/\sqrt2)(x_1-ix_2);\tilde z=(1/\sqrt 2)(x_3+ix_4)$ and $z=(1/\sqrt
2)(x_3-ix_4)$. The metric of signature $(4,0)$ and $(2+2)$
is, respectively, $ds^2=dyd\tilde y+(-)dzd\tilde z$ and the
complexified-spacetime  SDYM equations are
$F_{yz}=F_{\tilde y\tilde z} =0$ and $F_{y\tilde y}+(-)F_{z\tilde z}=0$.
Such equations require an $explicit$ signature-dependent definition of the
Hodge * operation which is consistent with the double-Wick rotation of the
Euclidean SDYM equations and with the fact that $^{**}F= F$. For our $choice$
of
coordinates : $\epsilon_{y\tilde y z \tilde z} =1$. The * operation in
spaces where $t=0,2$ is taken to be such that one of the SDYM eqs is  :
 $F_{y\tilde y} = i^t/2~\sqrt {|g|}\epsilon_{y\tilde y z \tilde
z}F^{z\tilde z}$;  where  $F^{z\tilde z}= -F_{z\tilde z}.$
The internal coordinates, $q,p$ can be incorporated into a pair of
$complex$-valued, canonical-conjugate variables ; $\hat q= Q(q,p).
{}~\hat p= P(p,q)$ such as $\{\hat q,\hat p\}_{qp}=1$. $Q,P$ are independent
maps from a sphere ( $S^2\sim CP^1$), per example, to $C^1$, such as
$Q\neq \lambda P;~\lambda =constant $.  This is in agreement with the
fact that the true symmetry algebra of Plebanski's equation is the $CP^1$
extension of the $sdiff~\Sigma$ Lie-algebra as discussed by Park [10].

    It was the suitable dimensional reduction in [1] :
$\partial_y=\partial_{\hat q};~ -\partial_{\tilde y}=\partial_{\hat p}$
and the ansatz (where for convenience we drop the ``hats'' over the $q,p$
variables ) :

$$\partial_zA_{\tilde z}=(1/2\kappa^2)\Omega,_{zq};
\partial_{\tilde z}A_z=(1/2\kappa^2)\Omega,_{\tilde z  p};\eqno(1a)$$
$$\partial_pA_{z}=(1/2\kappa^2)\Omega,_{pp};
\partial_qA_{\tilde z}=(1/2\kappa^2)\Omega,_{qq};\eqno(1b)$$
$$\partial_qA_{z}=(1/2\kappa^2)\Omega,_{pq};
\partial_pA_{\tilde z}=(1/2\kappa^2)\Omega,_{pq};\eqno(1c)$$
$$A_y=(1/2\kappa^2)\Omega,_q ;~~~A_{\tilde y} =-(1/2\kappa^2)\Omega,_p.
 \eqno(1d)$$
that yields the Plebanski equation in [1] .  $\kappa$ is a constant that has
dimensions of length and can be set to unity. The semicolon stands for partial
derivatives and $\Omega(z,\tilde z;\hat q,\hat p) $ is the  Plebanski's second
heavenly function. Upon such an anstaz and dimensional reduction, Plebanski's
second
heavenly equation was obtained in [1] :
$$(\Omega,_{\hat p\hat q})^2 -\Omega,_{\hat p\hat p}\Omega,_{\hat q\hat q}
+\Omega,_{z\hat q}
-\Omega,_{\tilde z  \hat p} =0. \eqno(2)$$

Eq-(2) yields self-dual solutions to the complexified-Einstein's equations,
and gives rise to hyper-Kahler metrics on the complexification of $T^*\Sigma$,
through a continuous self-dual deformation, represented by $\Omega$, of
the flat  metric in $(T^*\Sigma)^c$ [1].

  One of the plausible  first steps  in the dimensional-reduction of
Plebanski's equation is to
take a real-slice. A natural $real$ slice can be taken
by setting : $\tilde z=\bar z.~\tilde y=\bar y$ which implies, after using
: $\partial_y=\partial_{\hat q} ;~-\partial_{\tilde y}=\partial_{\hat p}$, that
$-(\partial_{\hat q})^*=\partial_{\hat p}$ and, hence, the Poisson-bracket
degenerates to zero; i.e. it ``collapses'' :
The quantity :$\{Q,P\}_{q,p}=\{Q,-Q^*\}_{q,p}$, if real, cannot be equal to $1$
but is
zero as one can verify by taking complex-conjugates on both sides of the
equation.  Therefore, since  the Poisson brackets between any two
potentials ,
$\{A_1 ,A_2\}_{qp}=\{A,B\}_{Q,P}\{Q,P\}_{pq}=0$,  the $CP^1$-extension of the
 $sdiff~\Sigma
$ Lie-algebra is $Abelianized$ ( no commutators) in the process.   DKS [3]
already made the
 remark that their derivation was  also valid for
$U(1)$. To sum up, taking a $real$ slice  reduces  $C^4$-dependent solutions to
$C^2$-dependent  ones ``killing'', in the process, the Poisson-brackets.

     The reader might feel unhappy with this fact. Another option is
 $ not$ to  take a real slice but, instead, one
 imposes the $C^1$-valued
dimensional-reduction condition ( the complexification of eq-3c, below
) : $\partial_{x_1}-\partial_{x_3}=0$; where $x_1,x_3$ are $complex$
coordinates. Since in this case $Q^*$ is $ no$ longer equal to  $-P$, the
Poisson-bracket  is well defined, one  ends up still  having the
$CP^1$-extended $sdiff~\Sigma$ Lie-algebra untarnished  and with a
$C^3$-dependent theory ( since the Plebanski equation was dependent of $C^4$).
Complex-valued gauge potentials is precisely what is needed in order to have
the
$CP^1$-extended $sdiff~\Sigma$ to be locally isomorphic to $su^*(\infty)$.
Following  the same step by step procedure as the one outline below, yields
a $complexification$ of the KP equation [2].

   The second step of the dimensional-reduction (in the real case) is to take
$\partial_{x_1}-\partial_{x_3 }=\partial_{x_{-}}=0$;
$x_{-}= x_1 -x_3.$  and,
hence, we end up with an effective $real$ three-dimensional theory.

Now we are ready to establish the correspondence with the dimensionally-reduced
$sl(2,R)$ SDYM
equations in $(3+3)$ dimensions by DKS [3]. Set      :

$$x_1\rightarrow X_6=Y;~~x_2\rightarrow x_2(X,Y,T);~~x_3\rightarrow X_3=X;
{}~~x_4\rightarrow X_1=T. \eqno(3a)$$
$$\partial_{X_2}= 0.~~\partial_{X_3}=\partial_{X_4}.
{}~~\partial_{X_5}=0. \eqno(3b)$$
$$\partial_{x_{-}}=0 \Rightarrow u(X;Y;T) \rightarrow \Omega
(x_1+x_3;x_2;x_4)=\Omega (x_{+};x_2;x_4).\eqno (3c)$$
(Notice the variables in eq-(3c); $\Omega$ is a function of a spatial, timelike
and $null$ variable. Compare this with the variables in the KP
function ; two temporal  and one spacelike. For us, $X_2,X_4,X_6$ are
spacelike and $X_1,X_3,X_5$ timelike.)
 the Jacobian :
$$     {\cal J}=\frac {\partial (X,Y,T)}{\partial (x_+,x_2,x_4)}=1
 \eqno (3e) $$

    Eq-(3a) defines a class of maps from ${\cal M}^{1+2}\rightarrow {\cal
N}^{1+1+1}$; i.e. $\phi : P(X,Y,T)\rightarrow P'(x_+,x_2,x_4)$. The
Jacobian $\cal J$ should not vanish and without loss of generality can be
set to one : ${\cal J}={\cal J}^{-1}=1 $. Hence, eq-(3a) defines a class of
volume-preserving-diffs.

     Using eqs-(3a-3c) in  equations (25,37,39,40,41) of DKS, we learn, from
the correspondence given in
eqs-(1,2) of [2], respectively, that a $one$ to $one$ correspondence with the
$SU(\infty)$  SDYM equations,
is possible iff we take for an Ansatz (see eq-46 in DKS)
$A_{x_1}=A_{x_3}.$ (eq-3d in [2]).  The DKS-Plebanski 'dictionary' reads :

$$F_{36}=0\rightarrow F_{x_1 x_3}=F_{y\bar z}+F_{\bar y z}=0
       . \eqno(4a) $$
$$F_{13} =0 \rightarrow F_{x_3 x_4}=iF_{z\bar z}=0 . \eqno(4b)$$.
$$F_{16}=0\rightarrow F_{x_1x_4}=F_{y\bar z}-F_{\bar y z}=0 . \eqno(4c)$$

The reason the right hand side of (4a) is zero is a $result$ of the
eqs-(3c,3d). Exactly the same happens to eq-(4b) which is nothing but one of
the  SDYM equations. $F_{y\tilde  y}= \{\Omega_{\hat q},\Omega_{\hat p}\}= 0$
is a result of the ansatz in
eqs-(1a-1d); the dim-reduction conditions in [1] and $all$  of the  $2+2$ SDYM
equations.
(This is $not$ the case in the Euclidean regime. After all there must be
subtle differences between the Euclidean versus the $2+2$ version; this is
one of them besides the fact that two timelike directions provide with two
Hamiltonian structures which are essential in many integrable systems ).
 Eq-(4b) becomes  after setting $\{\Omega_{\hat q},\Omega_{\hat p}\}=0$
in (2) :
$\Omega_{z\hat q}-\Omega_{\tilde z \hat p}=0$.
Equation-(4c) is zero iff (i). The ansatz of eq-(3d) in [2]  is used. (ii). The
dim-reduction conditions in eq-(3c) are taken and, (iii). The  eqs-(2,4b)
are satisfied. If conditions (i-iii) are met it is straight
forward to verify that eq-(3c) =$(\partial_{x_1}-\partial_{x_3})A_{x_4}=0$.

The $crux$ of [2] was to obtain the desired relationship between $u$ and
$\Omega$
in order to have self-consistent loop arguments and equations; to render
the right handsides of eq-(4a,4b,4c) to zero; and, to
finally, obtain the desired KP equation from the dimensional reduction of
the Plebanski equation (DRPE).

Having
estalished the suitable correspondence and the assurance that the right-
hand-sides of eqs-(4a,4b,4c) are in fact zero, we can now claim, by $cons
truction$, that
 equations-(39,40,41,47,53) of DKS [3] are the equivalent, in the $u$ variable
language,
to eqs-(4a,4b,4c) above, in the $\Omega$ language.
 Therefore the equivalence for eq-(4c) reads :

$$u_{T}-1/4u_{XXX}-3/2uu_{X}+(\lambda^2+\alpha\beta)u_{X}=$$
$$1/2(\partial {x^2_+}-\partial {x^2_2}
-i\partial {x_+}\partial {x_4}
-i\partial {x_+}\partial {x_2})\Omega
(x_{+};x_2 ;x_4 ) =0.  \eqno (5a)      $$

And, similarly, eq-(47) of DKS is the equivalent of eq-(4a) above :

$$\beta u_{X}-
u_{Y}=1/2(-\partial^2_{x_+} +\partial_{x_2}\partial_{x_4}+
2i\partial_{x_+}\partial_{x_2})\Omega =0.\eqno(5b)$$

and we include the DRPE  :

$$1/2(\partial^2_{x_+} + \partial_{x_2}\partial_{x_4})\Omega (x_{+};x_2;x_4)=0
\rightarrow (DRPE)\eqno (5c)$$

The function, $u (X,Y,T)$, satisfies the KP equation :

$$\partial_{X}(u_{T}-1/4~u_{XXX}-3/2~uu_{X})=-(\lambda^2+\alpha\beta)\beta^{-2}~u_{YY}.\eqno (6)
)$$
after  using the relation, $u_{Y}=\beta u_{X}\rightarrow
 u_{YY}=\beta^2 u_{XX}$ and differentiating the l.h.s. of (5a).
 One has to make a suitable scaling of the variables because the
KP equation obtained by DKS was given in terms of dimensionless quantities.
We must emphasize that $\Omega$ is not constrained, in any way whatsoever,
to satisfy two independent differential equations. The l.h.s of (5a) is
zero as a consequence of (4a,4b). Viceversa, if (4a,4c) are satisfied then
(4b) follows. We have seven equations to deal
with. These are :

(i). The lhs and rhs of eqs (5a); (ii). The lhs and rhs of eqs (5b);
(iii). The DRPE, eq-(5c); (iv). The KP equation, (6); and, (v). Equation
(3e), nonvanishing Jacobian.

   This is  all what is needed in order to  arrive finally to our main result
of [2] :
For every solution  $\Omega$ of the DRPE (5c) we set : $\Omega [x_+(X,Y,T);x_2
(X,Y,T);x_4(X,Y,T)]=u(X,Y,T)$ and plugging $\Omega$  into the $l.h.s$ of
(5a,5b) we
get three partial differential equations, once we include (3e) :
 ${\cal J}={\cal J}^{-1}= 1$, for the volume-preserving diffs, $\phi :
P(X,Y,T)\rightarrow P'(x_+,x_2,x_4)$. Once a solution for the three
diffeomorphisms that comprise $\phi$ is found,  then we have an explicit
expression for
$u(X,Y,T)$ that solves the KP equation $by~construction$.

And, viceversa, once a solution for the KP equation is found  we
set  $u[X(x_+,x_2,x_4);Y();T()]$
equal to $\Omega (x_+,x_2,x_4)$ and plugging
$u$ into the $r.h.s$ of (5a,5b) we get three partial differential equations
, once we include (3e) (nonvanishing Jacobian), for the inverse
volume-preserving-diffs, $\phi^{-1} :
P'(x_+,x_2,x_4)\rightarrow P(X,Y,T)$. Once a solution is found, we then
have an explicit expression for $\Omega (x_+,x_2,x_4)$ that solves the DRPE
$by~construction$ because eqs-(4a,4c) $\Rightarrow$ eq-(4b).

   Now, we arrive at the most important result of [2]. This
construction that relates the KP to Plebanski is precisely what furnishes
the geometrical meaning of $W$ gravity. Remember what we said earlier about
the fact that the  DRPE provides a solution-space, $\cal S$,  of dim-reduced
hyper-Kahler metrics in  the
$complexified$ cotangent space of the Riemannian surface, $(T^*\Sigma)^c$,
required in the
formulation of $SU^*(\infty)$ $2+2$ SDYM theory. Since the
volume-preserving diffs, $\phi;\phi^{-1}$, yield the dim-reduced- Plebanski-
KP correspondence; it is natural that the $W_{\infty}$ symmetry algebra
associated with the KP equation $is$ the $\phi~transform $
of the $dimensional~reduced~CP^1$-extended $sdiff~\Sigma$- Lie algebra.

   Because these $\phi$, volume-preserving- diffs, act on the $ solution~
space, \cal S$, alluded earlier, it becomes clear that the $W~metric$
can be interpreted as the gauge field which gauges these $\phi$
diffeomorphisms acting on the space $\cal S$ !!
In [1] we already made the remark that one could generalized matters
evenfurther by starting with a $SU(\infty)$ SDYM theory on a
$self~dual~curved~2+2$
background. Upon imposing the ansatz of eqs-(1a-1d) one gets a
generalization of equation (2) where an extra field, $\Omega^1$, the
Plebanski first heavenly form, appears in addition to $\Omega$. $\Omega^1$
is the field which generates  the self-dual metric for the $D=2+2$ background.
The generalization of (2) encodes the interplay between ``two'' self-dual
``gravities''; one stemming from $\Omega$, the other from $\Omega^1$. In
any case, in we wish to gauge the volume-preserving-diffs, $\phi$, in order
to get $W$ gravity  where the $W$ metric is the $gauge$ field, we need to
come up with an extra field, which is pressumably  the role played precisely by
$\Omega^1$.

   Upon attaching fibers at each ``point'', $\Omega$ of $\cal S$, these $\phi$
volume-preserving diffs  act as gauge transformations (automorphisms) along
the fiber;  since $\Omega$ remains fixed as we vary $\phi$ along the fiber,
one
generates  a continuous family of functions $u,u',u'',.....$ obeying the KP
equation. Therefore, the $\phi~ transform$ of the dim-reduced
$CP^1$-extended $sdiff~\Sigma$-Lie algebra is indeed the symmetry algebra
of the KP equation; i.e. A classical $W_{\infty}$ algebra which rotates
solutions of the KP equation into eachother. We speak loosely when we say
$W_{\infty}$ algebra; i.e; we also could be referring to the $W_{1+\infty}$
algebra. This all depends on the Topology of the two-surface, $\Sigma$. The
project for the future is to elucidate what types of $W_{\infty}$-algebras
one generates in this fashion. Our  aim is to have a $universal~W_{\infty}$
algebra from which $all~W_{\infty}$ algebras can be obtained by a judicious
truncation. Then one can pursue their quantization. All this is, of course,
a non-trivial task.

    What happens when we set the Jacobian, $\cal J$ to an arbitrary
constant, $\lambda$? In this case one gets a class of
volume-preserving-diffs depending on $\lambda,~\phi_{\lambda}$ which
furnishes the $W_{\infty}(\lambda)$ algebras discussed in the literature
, see references in [11]. A further and detailed discussion of this will be
presented in a
future publication.

   In the case that one imposes a $truncation$ the algebra should reduce to
$W_N$. A very natural truncation  can be done by a filtering procedure :
$SU(N-1)\subset SU(N)\subset SU(N+1)\subset .....$. It was in this way
that Hoppe [6] showed that a $basis$ dependent limit of $SU(N)$
when $N \rightarrow \infty)$ is
isomorphic to the $sdiff~S^2$-Lie algebra.  A 'similar' construction
based on $Jet~Bundles$ over  Riemannian surfaces and a  Drinfeld-Sokolov
reduction was given  by Bilal, Fock and Kogan to explain the origins of $W$
algebras [12]. A thorough discussion on
area-preserving diffs  and (classical) $w_{\infty}$ algebras was given by
Sezgin [6]. The Topology of $\Sigma$ is crucial as to what type of
$ w_{\infty}, w_{1+\infty}$ algebras one gets. For higher $genus$
surfaces, $g>1$ one can start with Siegel's upper-half of the complex-plane
( a disc, whose $sdiff$ Lie algebra is isomorphic to $SL(\infty)$) and take
suitable quotients by finite/discrete groups. This won't concern us at the
moment. There is a large  mathematical literature on this subject. What
mostly  concerns us is what are the physical implications to string theory
 and how to select its true vacuum state. This will see shortly.

      We've been working with a dim-reduction procedure; is it possible to
incorporate other types of reduction schemes, like Killing symmetry types ?
 Park [10] showed  that a Killing symmetry reduction of the Plebanski first
heavenly equation yields the $sl(\infty)$ continual Toda equation whose
asymptotic expansion by Ablowitz and Chakravarty [8] led to the KP
equation.  The (classical) $W_{\infty}$ algebra was obtained as a Killing
symmetry
reduction of the $CP^1$-extended $sdiff~\Sigma$-Lie algebra.  Since the physics
of self dual gravity  should be  independent
of which formulation one is using, either in terms of the first or second
heavenly form, this fact corroborates, once more,  that our findings should be
correct because
$sl(\infty,C)$ can be embedded into $sl(\infty,H)\sim su^*(\infty)$.
Gervais and Matsuo [13] proposed viewing $W$ gravity as holomorphic  embeddings
of $\Sigma$ into higher-dim Kahler manifolds, like $CP^n$ or  some suitable
coset space, $G/H$, where $W$ transformations look like particular diffs in the
target space.
The $W$ surfaces in $CP^n$ were shown to be instantons of the corresponding
non-linear $\sigma$-models. A geometrical meaning of the associated
conformally-reduced WZNW models was given in terms of the Frenet-Serret
equations for these $CP^n$ embeddings. The relevance of these  WZNW models
was already pointed out by Park [10] who showed  that WZNW models valued in
$sdiff~\Sigma$ were equivalent to $D=4$ SD gravity. In [14] we found  that the
$N=2$ SWZNW model valued in $sdiff~\Sigma$ was SDSG in four dimensions.
We see once again the crucial role that SDG and SDSG  has on
the nature of the (super) $W$ geometry. All these findings
 would fit  into our picture when a Killing-symmetry reduction scheme is
chosen.

   Another clue that we are on the right track was
provided by Boer and Goeree [15] who studied arbitrary $W$ algebras related
to the embeddings of $sl(2)$ in a Lie algebra $\cal G$. A simple general
formula for all $W$ transformations enabled them to construct covariant
actions for $W$ gravity and this action was nothing but a Fourier transform
of the WZNW action. It was precisely such general formula that provided a
'geometrical' interpretation of $W$ transformations as a homotopy
contractions of ordinary gauge transformations. These homotopy-contractions
should correspond  to
the volume-preserving $\phi$-diffs alluded earlier. For further geometrical
interpretations of $W$ transformations see Figueroa-O'Farril et al [16]; Yi
and Soloviev [17]; Sotkov et al [18].  Giveon and Shapere [19] noticed the
importance of volume-preserving diffs in the gauge symmetries of the $N=2$
string. Since it is their physics we are mainly concerned with, it is the
geometry of $N=2$ strings that we turn our attention to in
the conclusion of this  letter.

\section {Conclusion : The SKP Equation and the Geometry of $N=2$ Strings}

 Using the results by us in [1] we
can automatically extend the construction here to the supersymmetric case.
The Lorentzian version of the Plebanski equation was derived for
$(3+1)$ superspace :
$$(\Theta,_{\hat p\hat q})^2 -\Theta,_{\hat q\hat q}\Theta,_{\hat p\hat p}+
\Theta,_{\hat qz}-\Theta,_{\hat p\tilde  z}=0.   \eqno (7)$$

where $\Theta (z,\tilde z, \hat q,\hat p;\theta^+)$ is a light-cone chiral
superfield described by Gilson et al
[20]. The derivatives in (7) are taken with respect to the
left-handed variables variables : $ x^m \equiv x^m +i\theta \sigma^m \bar
{\theta}$. For details and  self-duality conditions in Euclidean and
Atiyah-Ward spacetimes,
spaces of signature $(2+2)$,  see [1,4,14,20].  A double-Wick rotation from
$2+2$ ( or a single one from  $3+1$) SDSG into Euclidean SG can only be
performed iff $N=even$ because
there are no Majorana spinors in Euclidean space. A simple Wick-rotation of
$3+1$ SDSG  into $2+2$ SDSG  should be possible, after a suitable
 truncation, since SDSG in $2+2$-dim is based on the existence of
Majorana-Weyl spinors.  In any case
it is not difficult to see due to the fact that we were working in
$complexified$ superspaces [21] (these authors also discussed quaternionic
superspaces),  that a  $N=1,~2+2 SU^*(\infty)$
complexified  SDSYM theory
accomodates a  doubling of the number of real degrees of freedom.
Therefore, one should get the $N=2$ SKP equation and its associated $N=2$
super-$W_{\infty}$ algebra. Nishino
has conjectured recently  the existence of  $N=2$ SKP equations in [4]. If we
had started initially  by
taking a $real$ slice,  a $N=1$ SKP equation would have been obtained.
It  is not difficult to see that in order  to embed the $N=4~2+2$ SDSYM theory
[4], one must start from a
$N=1~3+3~SU^*(\infty)$ complexified SDSYM theory instead; a
complexified-fiber-bundle over
a complexified six-dim superspace.  ( we are not concerned with the mathe
matical
subtleties for the time being, like the non uniqueness of the $N\
rightarrow$ ${\infty}$ limits; the existence of complex
structures in infinite-dimensional spaces;..... ).  A  dim-reduction to $2+2$
dimensions yields  a doubling,  which in turn, implies a fourth-folding
in terms of  the number of  real degrees of freedom (supersymmetries) .
 Therefore, for those algebras which can be embedded into
$su^*(\infty)$, we have that $six$ dimensional $N=1~3+3~SU^*(\infty)$
complexified SDSYM {\bf is} the $''master''$ theory of lower-dim-
supersymmetric integrable systems!
One could have started from an $N=1~D=5+5~SYM$ coupled to $N=1~D=5+5~SG$ and
after  suitable dimensional reductions/ truncations
obtain different types of $N$-extended lower-dim supersymmetric theories. A
further
constraint would extract the Self-Dual pieces from the latter [4]. However,
this
is not necessary !  We have now  at our disposal ways to generalize the notion
of
self-duality to higher than-four dimensions [23]. It $is$ the moduli of these
higher dimensional ($3+3$) $SU^*(\infty)$  $complexified$
generalized-self-dual super
symmetric theories that encode the geometry of $N=2$
strings. This is the main result of this letter.

     In view of this we can see that the geometry of $N=2$ open
superstrings is tightly connected to that of the moduli space of
$N=1~3+3~SU^*(\infty)$ complexified SDSYM theory. The super $W_{\infty}$
symmetry  would follow by extending our bosonic-construction in [2] to the SKP
equation. Surely enough, the latter moduli space (and the  associated with the
closed
(heterotic) $N=2$ superstring)  must be related to the
infinite-dimensional super-Grassmannian where super-Riemannian surfaces
(SRS) of arbitrary genus
are represented as 'points'. We learnt  from the
Manin-Radul SKP hierachy [24]  that the super-$tau$ function  satisfies the
super-Hirota equations of the SKP hierachy which  is the generating function
for the super-conformal field theory on the SRS of
arbitrary genus. An infinite perturbation expansion, in principle, should
allow  us to extract $nonperturbative$ information to our superstring theory.
And where, if any, should  this nonperturbative information come from ??
 From the moduli space of $D=2+2~SU^*(\infty)$ Super-Yang-Mills instantons.
We should not be worried about using complex-valued gauge-potentials as
long as we know how to extract real-valued information from the theory. We
must not forget that the chiral counterparts, ${\bar W}$ algebras,  are
always there!: the $CP^1$-extended $sdiff~\Sigma$-Lie algebra $does$
contain both the $W,\bar W$ algebras [10] .  Therefore,
we expect to see  a connection between the large $N$ limit of $SU(N)$
complexified Gauge Theories
and  the infinite genus-limit of Riemann surfaces in the sum over
world-sheets approach to string theory . Based on the counting above and on
the results in [1,2],  starting  with a
$N=4~SU^*(\infty)$ complexified SDSYM theory coupled to a  $N=4~2+2~SDSG$
complexified-background should
establish a link with Siegel's observation that $N=8$ (gauged) SDSG in $2+2$
is the underlying theory of the (heterotic) closed $N=2$ superstring.

    Since  $W_{\infty}$ is a  symmetry of  string-field theory,
[25], the quest for the true vacuum of string theory might, in principle,
not be so far away because, at least, we know where to look:
It ought to  lie  behind the moduli space of  $SU^*(\infty)$  SDSYM in $3+3$
dimensions. This might also  shed  some light as to how
 extract $D=4~ QCD$ from string theory in four dimensions. All this has to
be proven, of course. What we know for certain is that the construction of [2]
did not
require any laborious nor sophisticated algebra; it is right there in
section $II$!  Supersymmetric theories in $D=2+2$ have been constructed
in [4]; generalized-self-dual theories in [23]; therefore, the tools appear to
be  available to tackle and prove/disprove our
conjecture.

\vspace{7 mm}

\centerline {REFERENCES}
\vspace{7 mm}

1. C. Castro : Journ. Math. Phys. {\bf 34}, 2  (1993) 681.

2. C. Castro :'' The KP equation from Plebanski and $SU(\infty)$ SDYM theory''

   I.A.E.C-7-93 preprint. Submitted to the Jour. Math. Physics.

3.  A.Das, Z. Khviengia, E. Sezgin : Phys. Letters {\bf B 289} (1992) 347.

4. H. Nishino : UMDPP-93-preprint :'' Supersymmetric -KP systems Embedded

   in Supersymmetric Self-Dual Yang-Mills Theory. UMDEPP-93-145.

   ``SDSYM Generates Witten's Topological Field Theory'' UMDEPP-93-14.

    S.J.Ketov, S.J.Gates Jr. and H. Nishino : Nucl. Phys. {\bf B 393} (1993)
149.

5. W. Siegel : Stony Brook ITP-SB-92-24 (May 1992). H. Ooguri and C. Vafa.

   Nucl. Phys. {\bf B 361} (1991) 469; {\bf  ibid. 367} (1991) 83.

6. Jens Hoppe : Int. Journal Mod. Phys. {\bf A4} (1989) 5235.  E. Sezgin :

   ``Area-Preserving Diffeomorphisms , $w_{\infty}$ Algeb  and $w_{\infty}$
   Gravity. CTP-TAMU-13-92.

7.   J.Barcelos-Neto, A. Das, S.Panda and S. Roy. Phys. Lett. {\bf B 282}
(1992) 365.

     Cambridge University Press. (1991).

8. M.J. Ablowitz, P.A. Clarkson :{\bf ``Solitons, Nonlinear Evolution Equations

    and  Inverse Scattering''} London Math. Soc. Lecture Notes {\bf 149};

   Cambridge University Press (1991).

9. Edward Witten :''Surprises with Topological Field Theories ``

   Proceedings of Strings' 90 at College Station, Texas, USA.

10.  Q.H. Park : Phys. Letters. {\bf B 236} (1990) 429; {\bf ibid  B 238}
(1990) 287.

   Int.Jornal of Modern Phys {\bf  A 7}; page 1415, (1991).

11.  P. Bouwknegt, K. Schouetens : ``W symmetry in Conformal Field Theories

`` CERN-TH-6583/92. To appear in Physics Reports.

12.  A.Bilal, V.V. Fock and I.I. Kogan : ``On the Origins of $W$ Algebras.

   CERN-TH-5965-90 preprint.

13. J.L.Gervais, Y. Matsuo :'' Classical  $A_n~W$ geometry'' L.P.T.E.N.S

    91-35 preprint. Phys. Lett.{\bf  B 274} (1992) 309-316.

14. C. Castro : ``The $N=2~ SWZNW$ model valued in $ sdiff~\Sigma$  is

    Self Dual Supergravity in four Dimensions ``. I.A.E.C-11-92 preprint;

     submitted  to the Journ. Math. Phys.

15. J. de Boer and J. Goeree : ``Covariant $W$ Gravity and its Moduli Space

     from Gauge Theory'' THU-92-14 preprint. Univ. of Utrecht.

16. J. Figueroa-O'Farril,E. Ramos and  S. Stanciu : ``A Geometrical

     Interpretation of Classical $W$ Transformations. BONN-HE-92-27

17. W.S.Yi and O.A. Soloviev : `` $W_3$ gravity from affine maximal

    hypersurfaces''. UMDEPP-92-146.

18. G. Sotkov, M. Stanishkov. Nucl. Phys. B 356 (1991) 439.

19.  A. Giveon and A. Shapere : ``Gauge Symmetries of the $N=2$ String''

      IASSNS-HEP-92-14 preprint.

20.  G.R.Gilson, I.Martin, A. Restuccia and J.G. Taylor : Comm. Math.

    Physics {\bf 107} (1986) 377.

21. J. Lukierski and A. Nowicki; Ann. Phys. {\bf 166} (1986) 164.

22.  E.G.Floratos, J. Iliopoulos, G.Tiktopoulos : Phys. Lett. {\bf B 217}
(1989)  285.

23. A.D. Popov; JINR Rapid communications, {\bf no.6}  57 (1992) 57.

S. Salamon; Invent. Math. {\bf 67} (1982) 143. R.S. Ward; Nucl. Physics

{\bf B 236} (1984) 381.

A.S. Galperin, E. Ivanov, V. Ogievetsky and E. Sokatchev; Ann. Physics.

{\bf 185} (1988) 1 and 22.  A. D. Popov; Mod. Phys. Letters {\bf A 7}

     no. 23 (1992) 2077.

24. Y. Manin and A.O. Radul; Comm. Math. Physics. {\bf 98} (1985) 65.

25. M. Kaku;{\bf  Introduction to Superstrings.}  Springer-Verlag, 1988.

    M. Kaku; {\bf Strings, Conformal Field Theory and Topology,
 An Introduction.}

 Springer-Verlag, 1990.

\end{document}